\def\yesanswfig{y }
\message{ Do you want auto
insertion of figures using epsf.tex ?(y or n) }\read-1 to\answfig
\ifx\answfig\yesanswfig%
\message{Figures must be named fig1.eps, 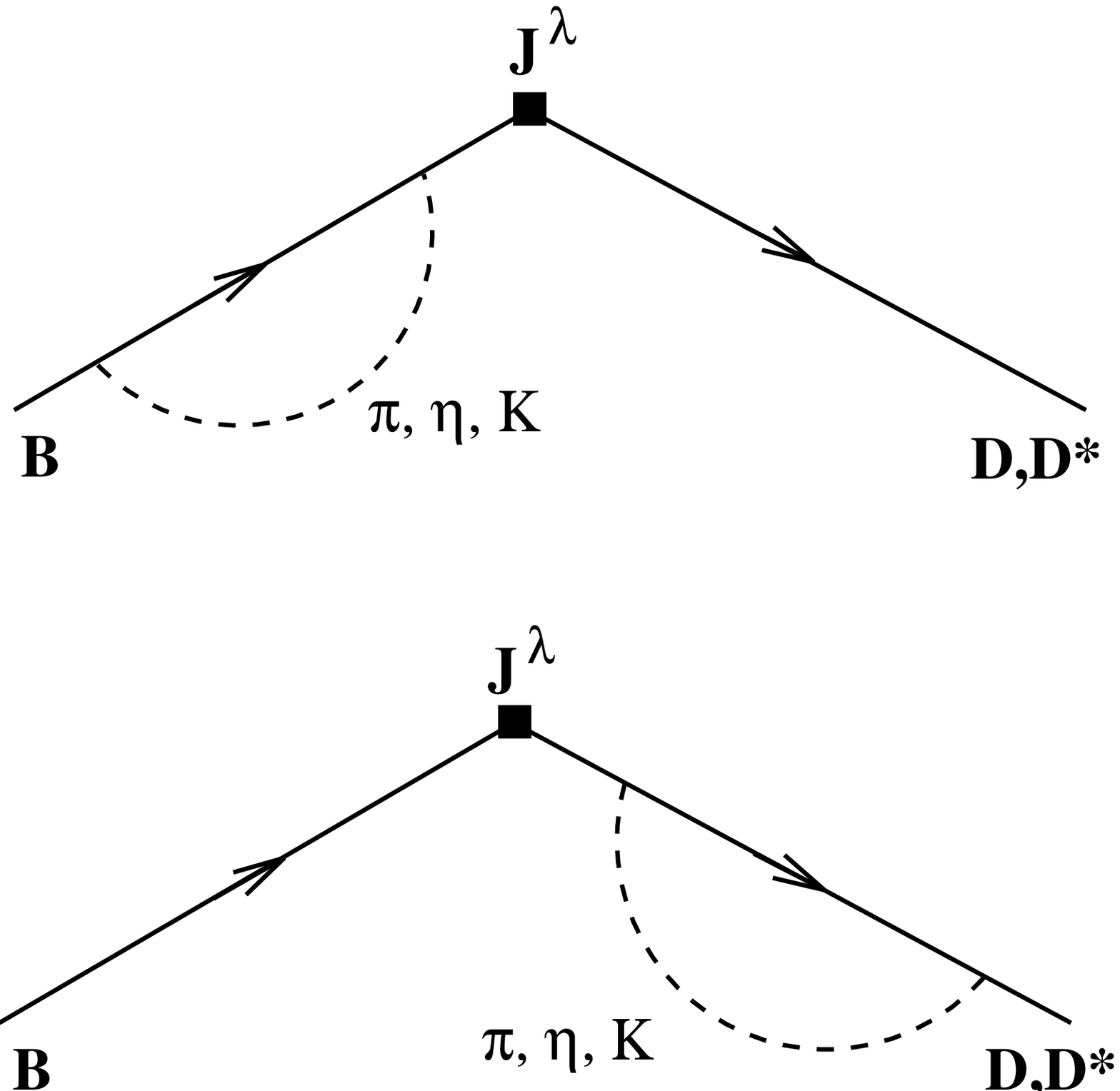,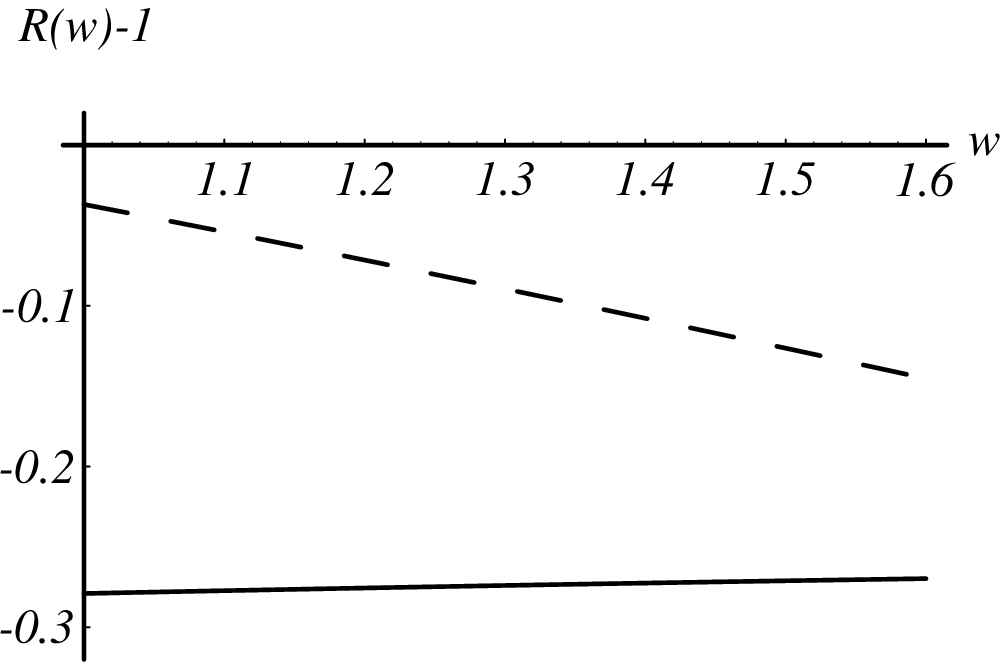,
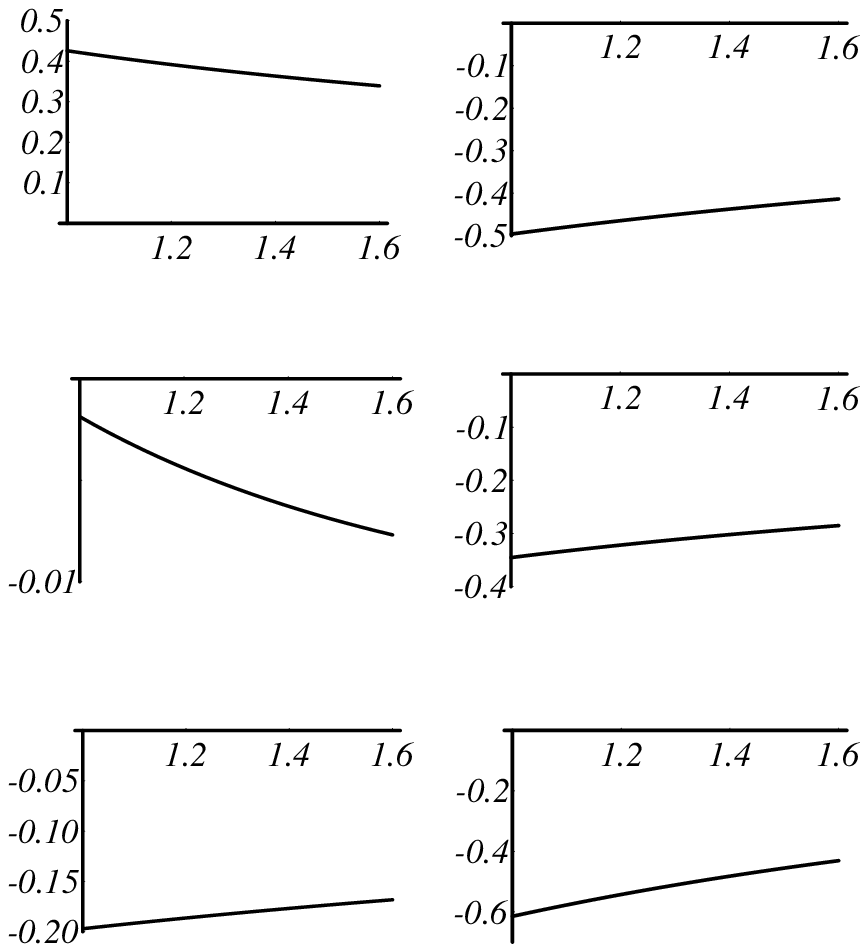,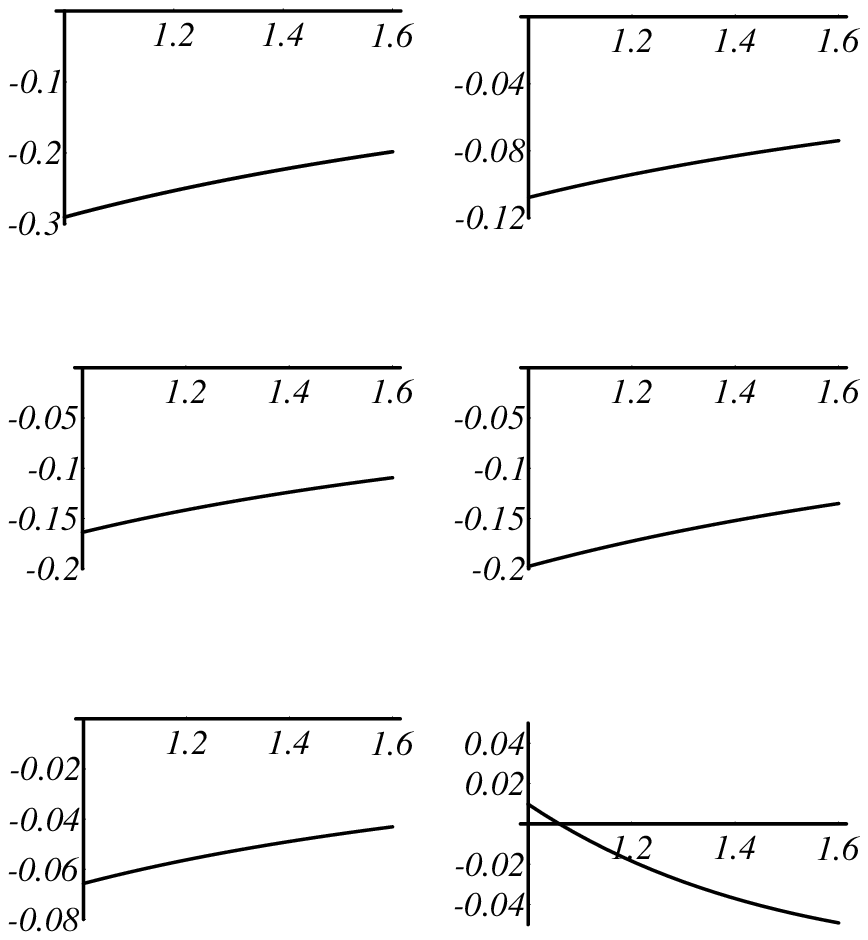,fig_c3.eps. 
They will be included automatically using epsf.tex}%
\input epsf
\def\INSERTFIG#1#2#3{\vbox{\vbox{\hfil\epsfbox{#1}\hfill}%
{\narrower\noindent%
\multiply\baselineskip by 3%
\divide\baselineskip by 4%
{\ninerm Figure #2 }{\ninesl #3 \medskip}}
}}%
\bigskip
\else\message{Figures will not be included automatically.
Figure captions will be appended at end}\def\INSERTFIG#1#2#3{}
\def\INSERTCAP#1{}
\fi
\message{We suggest replying `b' to the following:}
\input lanlmac
%

%from btok
\def\etal{{\it et al.}}

\def\ol{\overline}

\def\O{{\cal O}}

\def\H{{\cal H}}
\def\d{{\rm d}}
\def\im{{\rm i}}

\def\vslash{v\hskip-0.5em /}
\def\Aslash{A\hskip-0.5em /}

\def\gev{\;{\rm GeV}}
\def\OMIT#1{}
\def\frac#1#2{{#1\over#2}}
\def\smallfrac#1#2{{\textstyle{#1\over#2}}}
\def\eps{{\epsilon}}
\def\kintegral{\int{\d^{4-\eps}\,k\over(2\pi)^{4-\eps}}}

%my own

\def\ord{${\cal O}({1\over M}$)}

\def\lbar{\bar \Lambda}
\def\gt{\tilde g}

\def\vk{v \cdot k}

% my own for B_s -> D_s

\def\log{\ln \frac{m^2}{\mu^2} }
\def\xo{\xi^{{\vphantom{2}}}_0}
\def\xp{{\xi^{{\vphantom{2}}}_+}}
\def\xm{{\ol \Lambda \over 2} \xi_0}
\def\xmm{{\ol \Lambda } \xi_0}
\def\Xone{\chi^{{\vphantom{2}}}_1}
\def\Xtwo{\chi^{{\vphantom{2}}}_2}
\def\Xthree{\chi^{{\vphantom{2}}}_3}
\def\cone{\chi^{{\vphantom{2}}}_1}
\def\ctwo{\chi^{{\vphantom{2}}}_2}
\def\cthr{\chi^{{\vphantom{2}}}_3}
\def\C#1#2{C_#1^{(#2)} }
\def\D#1#2{D_#1^{(#2)} }
\def\vpk{v' \cdot k}
\def\w{\omega}
\def\Ha{H_a(v')}
\def\H{H(v')}
\def\Hb{\overline H(v)}
\def\Hbb{\overline H_b(v)}
\def\r#1{\rho_{#1}}

\def\cgb{ \overline c \gamma^\lambda (1 - \gamma_5) b}
\def\vlam{v^\lambda}
\def\vplam{{v'}^\lambda}
\def\vpD{v' \cdot \epsilon_D^*}
\def\vB{v \cdot \epsilon_B}
\def\epslam{\epsilon^\lambda_{\alpha\beta\gamma}}
\def\ordMm{$\CO(\frac1M,m_s)$}

\def\ordm{$\CO(\frac1{M^0},m_s)$}
\def\gB{{g^{{\vphantom{2}}}_B}}
\def\gD{{g^{{\vphantom{2}}}_D}}
\def\gDs{{g^{{\vphantom{2}}}_{D^*}}}
\def\gbd{ {\gB \gD \over 16 \pi^2 f^2} }
\def\gbds{ {\gB \gDs \over 16 \pi^2 f^2} }
\def\Dstring{ \(1- { 2 g_{D^*}^2 D_1^{(5D)} + g_D^2 D_1^{(3D)}
      +3 g_B^2 D_1^{(1B)} \over 32 \pi^2 f^2_K } \) }

% From rad
\def\np#1#2#3{{Nucl.~Phys.}~B{#1} (#2) #3}
\def\pl#1#2#3{{Phys.~Lett.}~{#1}B (#2) #3}

\def\physrev#1#2#3{{Phys.~Rev.}~{#1} (#2) #3}

\def\del{\partial}

\def\bar#1{\overline{#1}}

\def\bra#1{\left\langle #1\right|}
\def\ket#1{\left| #1\right\rangle}

\def\Tr{\mathop{\rm Tr}}

   %curly letters
  \def\CC{{\cal C}}

\def\CM{{\cal M}}  \def\CO{{\cal O}}

\def\[{\left[}
\def\]{\right]}
\def\({\left(}
\def\){\right)}

\def\sproj{{1\over 2} \(1+\vslash\)}

\def\gmu{\gamma_\mu}

\def\[{\left[}
\def\]{\right]}
\def\gamf{\gamma_5}

\Title{\vbox{\hbox{UCSD/PTH 94-26}\hbox{hep-ph/9502311}}}
{SU(3) Corrections to $B \to D l \overline \nu$ Form Factors at \ord}

\centerline{C. Glenn Boyd and Benjam\'\i n Grinstein}
{\it\centerline{Dept. of Physics 0319, University of
California at San Diego}
\centerline{9500 Gilman Dr, La Jolla, CA 92093}}

\vskip .3in
We compute the \ordMm\ heavy quark and SU(3) corrections to $\overline
B_s \to D_s e \nu$ form factors.  In the limit of vanishing light
quark mass, $\overline B_s \to D_s e \nu$ form factors are given in
terms of the the $\overline B \to D e \nu$ form factors, the leading
order chiral parameter $g$, and two \ord\ chiral parameters $g_1$ and
$g_2$.  All the chiral parameters can be extracted, in principle, from
other heavy meson decays. Analytic counterterms proportional to the
strange quark mass are presented for completeness, but no predictive
power remains when they are included. Anomalously large loop
corrections warn of poor convergence of the heavy quark chiral
symmetry expansion for these processes. This suggests that naive
extrapolations of $\overline B \to D$ form factors relying on heavy
quark and chiral symmetries, as often used in monte carlo simulations
of lattice QCD, may incur large errors.
\bigskip
\bigskip
%\centerline{DRAFT: NOT FOR PUBLICATION}
 
%\draft
\Date{February 1995}

\newsec{ Introduction}
 
In the standard model of electroweak interactions, the semileptonic
decays $\overline B_q\to D_q e \bar\nu$ and $\overline B_q\to D_q^* e
\bar\nu$ proceed via exchange of a charged vector boson and therefore
the corresponding rates are given in terms of the form factors for the
charged currents times the fundamental CKM parameter $|V_{cb}|$. The
study of these form factors is of interest to those who may try to
extract $|V_{cb}|$ from experimental measurements of the decay rates
as well as to those who are interested in how well approximate
chiral and heavy-quark symmetries work in nature.

The combination of chiral $SU(3)$ symmetry and heavy quark flavor-spin
symmetries gives all 18 form factors of the charged current
$J_\mu=\bar c\gamma_\mu(1-\gamma_5)b$ matrix elements between a $B_q$
meson and a $D_q$ or a $D_q^*$ mesons, with $q=u,d,s$, in terms of a
single `Isgur-Wise' function. Because these are only approximate
symmetries, the relations among form factors are not expected to hold
exactly. SU(3) chiral log corrections to the relations between form
factors have been examined at $\CO(M^0)$\ref\savjen{E. Jenkins and
M.J. Savage, \pl{281}{1992}{331}\semi J. Goity,
\physrev{D46}{1992}{3929} }. $\CO(\frac1{M^n})$ corrections involving
hyperfine splitting and inverse powers of the pion mass have also been
studied\ref\chow{C-K. Chow, M.B. Wise, \physrev{D48}{1993}{5202}
(hep-ph/9305229)\semi L. Randall and M.  Wise,
\pl{B303}{1993}{135}}. 

Here we present the first complete analysis of all \ord, $SU(3)$
breaking corrections to the relations between $B_q \to \{D_q,D^*_q\}$
form factors. In addition to hyperfine corrections, we include the
leading heavy quark symmetry breaking corrections in both the current
and the lagrangian. We account both for terms that depend
non-analytically on the symmetry breaking parameters $1/M$ and $m_s$,
and are enhanced in the theoretical limit of small $1/M$ and $m_s$,
and for terms with analytic dependence, which although suppressed in
the chiral limit are often non-negligible in reality.

If the analytic counter-terms are indeed non-negligible, predictive
power is lost. Of course, if they are small at some preordained
renormalization point, then the dominant non-analytic terms can be
calculated. We find that these non-analytic terms can be
substantial. In fact, simultaneous violations of both chiral and
heavy quark symmetries can be as large as 30\%.

Determinations of form factors for $\overline B\to D l\nu$ by Monte
Carlo simulations of lattice QCD often extrapolate in heavy and light
masses to the physical case\ref\latt{C. Bernard, Y. Shen, and A. Soni,
\np{~Proc. Suppl. 30}{1993}{473}\semi The UKQCD Collaboration,
\np{~Proc. Suppl. 34}{1994}{486}}. Our study indicates that such
extrapolations may incur large errors due to violations of heavy quark
and chiral symmetries.

To investigate $B\to D$, we write a chiral lagrangian that describes
the low energy interactions of single heavy mesons with pions. In this
lagrangian all of the symmetries are explicitly realized. We extend
the lagrangian to include symmetry breaking terms to the order of
interest, and retain only those terms that contribute to the
$B\to\{D,D^*\}$ form factors. As we will see, these corrections
involve heavy quark spin and flavor violating axial couplings $g_1,
g_2$, which may be extracted, in principle, from heavy to light meson
decays. An analogous analysis is done for the charged current
operator, which is written in terms of the meson fields.
%Analytic counter-terms to the current introduce unknown functions (form
%factors), but in the end there are still many predicted relations
%between form factors. 
One loop diagrams must be computed to complete
the calculations, since they give terms of the same order in the
expansion parameters $1/M$ and $m_s$ (or sometimes even lower order,
when they give non-analytic dependence on these parameters).

In section 2 we review the formulation of chiral lagrangians for
interactions of heavy mesons with pions, and construct the lagrangian
and charged current to linear order in $1/M$ and $m_s$. In the following 
section we describe the one loop calculation and present our results as 
corrections to hadronic form factors. We verify in section~4 that these 
results are consistent with Luke's theorem\ref\luke{M.E. Luke,
\pl{B252}{1990}{447}}, then discuss physical implications in section~5. 
We find surprisingly large symmetry violations for a particular ratio of
form factors, casting doubt on the convergence of the heavy quark-chiral
expansion for these processes. We make concluding remarks in the final section. 

\newsec{ Lagrangian and Current}
The low momentum strong interactions of $B$ and $B^*$ (or $D$ and
$D^*$) mesons are governed by the chiral lagrangian
\ref\wbdy{M.~Wise, \physrev{D45}{1992}{2188}\semi
G.~Burdman and J.~F.~Donoghue, \pl{280}{1992}{287}\semi
T.~M.~Yan \etal, \physrev{D46}{1992}{1148}}
\eqn\original{\eqalign{
       {\cal L}&=
    -\Tr\left[\overline H_a(v)\im v\cdot D_{ba} H_b(v)\right]\cr
    &+ g\,\Tr\left[\overline H_a(v)H_b(v)\,\Aslash_{ba}\gamma_5\right]\,. }}
Operators suppressed by powers of the heavy meson mass $1/M_B$,
factors of a light quark mass $m_q$, or additional derivatives have
been omitted.  The field $\xi$ contains the octet of
pseudo-Nambu-Goldstone bosons
\eqn\pseudo{\xi=\exp\(i\Pi/f\),}
where
\eqn\cm{\Pi=\pmatrix{{\textstyle{1\over\sqrt{2}}}\pi^0+{\textstyle
    {1\over\sqrt{6}}}\eta&\pi^+&K^+\cr\pi^-
       &{\textstyle{-{1\over\sqrt{2}}}}\pi^0+
   {\textstyle{1\over\sqrt{6}}}\eta&K^0\cr K^-&\bar K^0&-{\textstyle\sqrt
                   {2\over 3}}\eta}.} 
The bosons couple to heavy fields through the covariant derivative and
the axial vector field,
\eqn\covax{
    \eqalign{&D_{ab}^\mu=\delta_{ab}\del^\mu+V_{ab}^\mu
    =\delta_{ab}\del^\mu+\half\left(\xi^\dagger
    \del^\mu\xi+\xi\del^\mu\xi^\dagger\right)_{ab}\,,\cr
    &A_{ab}^\mu={\im \over 2}\left(\xi^\dagger\del^\mu\xi
    -\xi\del^\mu\xi^\dagger\right)_{ab}
    =-{1\over f}\partial_\mu{\cal M}_{ab}+{\cal O}({\cal M}^3)\,.}}

The $B$ and $B^*$ heavy meson fields are incorporated into the 
$4 \times 4$ matrix $H_a$:
\eqn\defineh{\eqalign{H_a&=\sproj\[\ol B^{*\mu}_a\gmu - \ol B_a\gamf\],\cr
             \bar H_a&=\gamma^0 H_a^\dagger \gamma^0\,.}}
The four-velocity of the heavy meson is $v^\mu$, and the index $a$
runs over light quark flavor. The bar over $B$ will sometimes be
omitted for notational simplicity.

The lagrangian to $\CO(\frac1M,m_s)$ may be written\ref\US{C. G.Boyd
and B. Grinstein, 
%{\it Chiral And Heavy Quark Symmetry Violation In B Decays,\/} 
UCSD-PTH-93-46, SMU-HEP/94-03, SSCL-Preprint-532 (hep-ph/9402340),
to appear in Nucl. Phys. {\bf B}}
\eqn\lagrangem{
{\cal L}= -\Tr\left[\overline H_a(v)\im
v\cdot D_{ba} H_b(v)\right] + {\tilde g_{{\ol H}H} }\,\Tr\left[
\overline H_a(v)H_b(v)\,\Aslash_{ba}\gamma_5\right]\,}
where
$$\gt = \cases{ \gt_{B^*} = g + \frac1M (g_1 + g_2) &for $B^* B^*$
coupling,\cr \gt_B = g + \frac1M (g_1 - g_2) &for $B^* B$
coupling.\cr}$$ 
Any $SU(3)$ counterterms relevant to $B \to D l\ol
\nu$ at this order may be accounted for by using normalized fields and
physical masses in propagators.  Our propagators are ${\im \over 2(v
\cdot k + \frac34 \Delta)}$ for the $B$, ${-\im (g^{\mu\nu} - v^\mu
v^\nu) \over 2(v \cdot k - \frac14 \Delta)}$ for the $B^*$, ${\im
\over 2(v \cdot k - \delta + \frac34 \Delta)}$ for the $B_s$, and
${-\im (g^{\mu\nu} - v^\mu v^\nu) \over 2(v \cdot k -\delta - \frac14
\Delta)}$ for the $B^*_s$, where $\Delta = M_{B^*} - M_B$ is the
hyperfine mass splitting and $\delta = M_{D_s} - M_D = M_{B_s} - M_B +
\O(\frac{\lbar^2}M) $ is the SU(3) mass splitting.

The current may be parametrized as
    \eqn\currentm{\eqalign{
  J^\lambda &= \[ -\xo(\w) 
       + \r1(\w)( \frac1{M_D} + \frac1{M_B})\] \Tr[\Hb \Gamma \H] \cr
      &+ \r2(\w)\[\frac1{M_D}  \Tr[ \gamma_5 \Hb \gamma_5 \Gamma \H]
       + \frac1{M_B} \Tr[ \gamma_5 \Hb \Gamma \gamma_5 \H]\]\cr
      &+ \r3(\w)\[\frac1{M_D}  \Tr[ \gamma_{\mu} \Hb \gamma^\mu \Gamma \H] 
       + \frac1{M_B}  \Tr[ \gamma_{\mu} \Hb \Gamma \gamma^\mu \H ]\]\cr
      &+ \r4(\w)\[\frac1{M_D}  \Tr[ \sigma_{\mu\nu} \Hb \sigma^{\mu\nu} \Gamma \H]
       + \frac1{M_B} \Tr[ \sigma_{\mu\nu} \Hb \Gamma \sigma^{\mu\nu} \H]\]\cr
      &+ \r5(\w)\[\frac1{M_D}  \Tr[ \Hb \vslash' \Gamma \H] 
       + \frac1{M_B} \Tr[ \Hb \Gamma \vslash \H]\]\cr
     &+ \r6(\w) \[\frac1{M_D}  \Tr[ \gamma_{\mu} \Hb \gamma^\mu \vslash' \Gamma \H] 
       +\frac1{M_B} \Tr[ \gamma_{\mu} \Hb \Gamma \vslash \gamma^\mu \H] \]\cr }}
where $\w = v \cdot v'$ and $\Gamma = \gamma^\lambda (1 - \gamma_5)$.  
Time reversal invariance dictates that all the $\rho$'s be real.

If we take $M_B \to \infty$, and consider only $B_q\to\{D_q,D_q^*\}$
matrix elements,  the resulting form for the current is
valid to all orders in $\frac1{M_D}$. After all it involves seven
independent functions, but the matrix elements can be expressed in
terms of six form factors. However, at $\CO(\frac1{M_D})$, heavy quark
symmetry is broken in a specific fashion, and relations among the
rho's exist.  The simplest way to find these relations is to match
onto the heavy quark effective theory\ref\hqeft{N. Isgur and M. Wise,
\pl{232}{1989}{113}; \pl{237}{1990}{527}\semi
E. Eichten and B. Hill, \pl{234}{1990}{511} \semi 
H. Georgi, \pl{240}{1990}{447} \semi
A. Falk, B. Grinstein, and M. Luke, \np{357}{1991}{185}}, 
giving\luke,
\eqn\matching{\eqalign{
  \r1 &= -\xp  -\xm  -  \Xone + \Xtwo - 2 \Xthree\cr
  \r2 &= 0\cr
  \r3 &= -(1 + \w)\xp  -(1-\w) \xm  +  \w \Xtwo - 2 \Xthree\cr\
  \r4 &= - \Xthree \cr
  \r5 &= -\xp + \xm +  \Xtwo \cr
  \r6 &=  \Xtwo \cr }}
In the notation of \luke, $\xo$ is the leading order 
Isgur-Wise function, while $\xp,\cone,\ctwo$ and $\cthr$ are \ord\ corrections
satisfying $\cone(\w=1) = \cthr(\w=1) =0$, and $\lbar = M_B - M_b$.
 
At \ordm, the current contains $SU(3)$ violating terms proportional to
the light quark mass matrix $m_q = {\rm diag}[0,0,m_s]$. To leading order
in derivatives, the chiral symmetry breaking current is
\eqn\sucurrent{\eqalign{
J^\lambda_{(m)} &= -\xo {\eta_0 \over \Lambda_\chi} \Tr[\Hb_b \Gamma \H_a]
\CM^+_{ab} 
     + {\kappa_1 \over \Lambda_\chi}\Tr[\Hb_a \Gamma \H_a] \CM^+_{bb}\cr
    &+ {\kappa_2 \over \Lambda_\chi}\Tr[\Hb_b \Gamma \H_a \gamma_5] \CM^-_{ab}
 + {\kappa_3 \over \Lambda_\chi}\Tr[\Hb_a \Gamma \H_a \gamma_5] \CM^-_{bb}\cr}}
where $\CM^{\pm} = \frac12(\xi m_q \xi \pm \xi^\dagger m_q \xi^\dagger)_{ba}.$

To \ordMm, only operators linear in $m_q$ and inserted in tree graphs
are relevant, so we take $\xi \to 1$, $\CM^+ \to m_q$, and $\CM^- \to
0$. Terms with $\CM^+_{aa}$ are $SU(3)$ symmetric and can be absorbed
by redefinitions of the parameters $\r1$--$\r6$.  This
limits considerably the terms that need be considered: To 
$\CO(M^0,m_s)$ only $\eta_0$ enters, while to~\ordMm\ we
have
\eqn\jMm{\eqalign{
 J^\lambda_{(M+m)} &= m_{ab}^q\Big\{({-\xo \eta_0 \over \Lambda_\chi} + {\r1
\eta_1 \over M_D} )
                  \Tr[\Hbb \Gamma \Ha] 
        + \r2 {\eta_2 \over M_D} \Tr[ \gamma_5 \Hbb \gamma_5 \Gamma \Ha]\cr
     &+ \r3  {\eta_3 \over M_D}\Tr[ \gamma_{\mu} \Hbb \gamma^\mu \Gamma \Ha]
     + \r4 {\eta_4 \over M_D} 
             \Tr[ \sigma_{\mu\nu} \Hbb \sigma^{\mu\nu} \Gamma \Ha]\cr
    &+ \r5 {\eta_5 \over M_D} \Tr[ \Hbb \vslash' \Gamma \Ha]
  + \r6 {\eta_6 \over M_D} \Tr[ \gamma_{\mu} \Hbb \gamma^\mu \vslash' \Gamma
               \Ha]\Big\}\cr
           }}

Since we are ignoring up and down masses, the current in Eq.~\jMm\
only contributes to the $B_s\to \{D_s,D_s^*\}$ transitions. As in the
$B\to \{D,D^*\}$ case, we can match onto the heavy quark effective
theory. The $\eta_2$ term multiplies $\r2$, which vanishes. In
addition, the parameters satisfy
\eqn\sumatch{
 - \r3 \eta_3 + 2 \r4 \eta_4 +(1+\w)\r5 \eta_5 -\r6 \eta_6 
                   = \ol \Lambda_s \xi_0,}
where $\ol \Lambda_s = M_{B_s} - m_b = \ol \Lambda + \delta$.
  
\newsec{ Calculation and Results}
It is convenient to introduce form factors for $B^*\to\{D,D^*\}$
transitions in addition to those for $B\to\{D,D^*\}$. We define them
as follows:
\eqn\defff{\eqalign{
\smallfrac1{\sqrt{M_D M_{B^*}}}
   \bra{D(v)} \cgb &\ket{B^*(\eps_B,v')} = f_0 \vB \vlam
+ \im f_1 \epslam \eps_B^\alpha v^\beta v'^\gamma\cr
&+ f_2 \vB \vplam + f_3 \eps_B^\lambda \cr
\smallfrac1{\sqrt{M_D M_{B^*}}}
   \bra{D^*(\eps_D,v)} \cgb &\ket{B^*(\eps_B,v')} =
  f_4 \eps_B \cdot \eps_D^* \vlam + f_5 \eps_B \cdot \eps_D^* \vplam \cr
  &+ f_6 v \cdot \eps_B  v' \cdot \eps_D^* \vplam
   + f_7  v \cdot \eps_B  v' \cdot \eps_D^* \vlam \cr
   &+ f_8  v' \cdot \eps_D^* \eps_B^\lambda
   +  f_9   v \cdot \eps_B \eps_D^\lambda \cr
   &+ \im f_{10} \epslam \eps_B^\alpha \eps_D^{*\beta} v^\gamma
   + \im f_{11} \epslam \eps_B^\alpha \eps_D^{*\beta} v'^\gamma\cr
 &+ \im f_{12} (v \cdot \epslam \eps_B \eps_D^{*\alpha} v^\beta v'^\gamma
+\eps_{\alpha\beta\gamma\delta}\eps_B^\alpha \eps_D{*^\beta} v^\gamma v'^\delta
\vlam) \cr
   &+\im f_{13} (v' \cdot \eps_D^* \epslam \eps_B^{\alpha} v^\beta v'^\gamma
-\eps_{\alpha\beta\gamma\delta}\eps_B^\alpha \eps_D^{*\beta} v^\gamma v'^\delta
\vplam) \cr
\smallfrac1{\sqrt{M_D M_{B^*}}}
  \bra{D^*(\eps_D,v)} \cgb &\ket{B(v')} =
    h_1 \vpD \vplam + h_2 \vpD \vlam + h_3 \eps_D^{*\lambda} \cr
    &+ \im h_4 \epslam \eps_D^\alpha v^\beta v'^\gamma\cr
\smallfrac1{\sqrt{M_D M_{B^*}}}
   \bra{D(v)}\cgb &\ket{B(v')} =
    h_5 \vlam + h_6 \vplam \cr
       }}
Other vector to vector form factors are simply related to $f_{12}$ and
$f_{13}$ by the identity $ \epsilon^{a b c d }g^{\mu e} +
\epsilon^{e a b c }g^{\mu d} +\epsilon^{d e a b }g^{\mu c} +
\epsilon^{c d e a}g^{\mu b} +\epsilon^{b c d e}g^{\mu a} =0 $.
The Feynman rules for an insertion of the current are 
summarized in terms of these form factors in 
appendix~A. We do the computation by using the form factors
above in the one loop diagrams, replacing them by their values in
terms of the chiral parameters $\rho_i, g_i$ at the very end.  We then express
the results of the one loop computation as corrections to the
form factors $h_1$ through $h_6$ (in the notation of \chow, 
$h_1 = \tilde a_+ - \tilde a_-, h_2 = \tilde a_+ + \tilde a_-, 
h_3 = \tilde f, h_4 = \tilde g, h_5 = \tilde f_+ + \tilde f_-$,
and $h_6 = \tilde f_+ - \tilde f_-$). 

\INSERTFIG{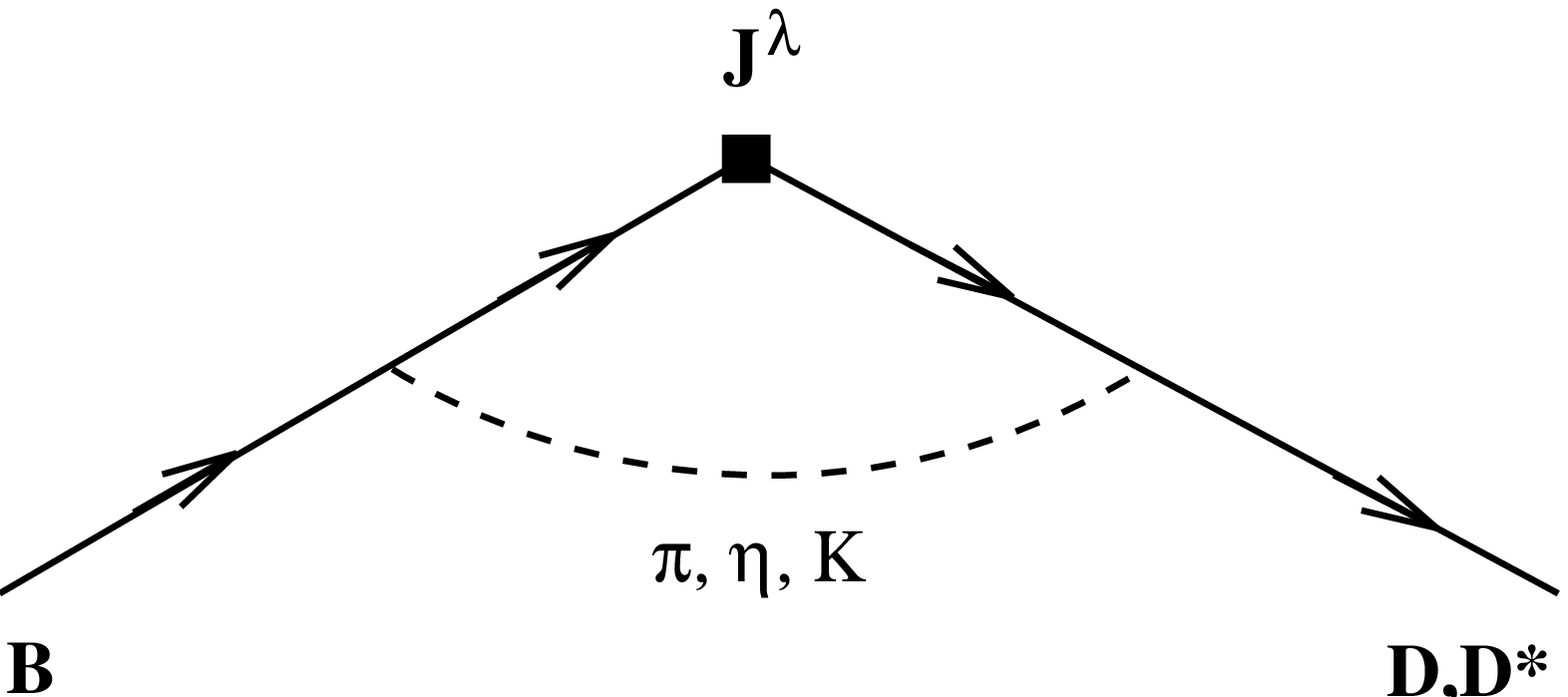}{1. }{Vertex correction graphs in the computation of
chiral corrections to form factors for $B\to\{D,D^*\}$ form factors.}

We turn our attention to the one loop corrections. The only diagrams
that need be computed are the vertex correction of \fig\vrtx{Vertex
correction graphs in the computation of chiral corrections to form
factors for $B\to\{D,D^*\}$ form factors.}\ and the self-energy graphs
of \fig\slfenrgy{Self-energy diagrams contributing to the chiral
corrections to form factors for $B\to\{D,D^*\}$.}.  The only integral
needed is
\eqn\Cdefin{\eqalign{
  C^{\alpha\beta}(\w,m,\Delta,\Delta') &= \kintegral { k^\alpha k^\beta
   \over  (k^2 - m^2)(\vk - \Delta)(\vpk - \Delta')  } \cr
    &= {\im \over 16 \pi^2} [C_1(\w,m,\Delta,\Delta') g^{\alpha\beta}
 + C_2(\w,m,\Delta,\Delta') ( v^\alpha v'^\beta + v^\beta v'^\alpha) ]\cr
 &+ C_3(\w,m,\Delta,\Delta')  v'^\alpha v'^\beta
    + C_4(\w,m,\Delta,\Delta') v^\alpha v^\beta \cr
                             } }
%\eqn\intgrlintrdcd{
%\kintegral { k^\alpha k^\beta
%   \over  (k^2 - m^2)(\vk - \Delta)(\vpk - \Delta')  } }
%
We express the four functions $C_i$ as one dimensional integrals in
Appendix~B. Also in appendix~B, we define and plot certain linear 
combinations,
$C_i^{(j)}(\w)$, of the functions $C_i(\w)$, and linear combinations
$D_i^{(j)}$ of the constants $C_i(1)$, that appear frequently in our
results.

\vfill\eject
   
\INSERTFIG{fig2.eps}{ 2.}{Self-energy diagrams corntributing to the chiral
corrections to form factors for $B\to\{D,D^*\}$.}
 
   Wavefunction renormalization factors are given by 
\eqn\wvfn{\eqalign{
   Z_B &= 1 - {3  g_B^2 \over 16 \pi^2 f^2}\D11 \cr         
   Z_{B_s} &= 1 - {3  g_B^2 \over 16 \pi^2 f^2}\D12 \cr         
   Z_{B^*} &= 1 - {1\over 16 \pi^2 f^2}(2 g_{B^*}^2 \D15
                                           + g_B^2 \D13) \cr         
   Z_{B_s^*} &= 1 - {1\over 16 \pi^2 f^2}(2 g_{B^*}^2 \D16
                                           + g_B^2 \D14) \cr  }}

Combining the results for vertex and self-energy graphs we find it now
straightforward to obtain the form factors valid to \ordMm\ at one
loop.  For decays to the pseudoscalar, we have
\eqn\hfive{\eqalign{
h_5^{B\to D}
 &= \(1 - {3 (g_B^2 D^{(1B)}_1
    + g_D^2 D^{(1D)}_1)\over 32 \pi^2 f^2}\) \Bigl\{\xo\cr
&\qquad \quad
+ \smallfrac1{M_D} \[ (2-\w)\xmm + 2(1+\w)\xp +\Xone +2 (1-\w)\Xtwo
    +6 \Xthree\] \cr
&\quad \qquad
+ \smallfrac1{M_B} \[-(2-\w)\xmm - 2(1+\w)\xp +\Xone +2 (1-\w)\Xtwo
    +6 \Xthree\] \Bigr\} \cr
&-{\gB \gD \over 16 \pi^2 f^2} C^{(1)}_1 \Bigl\{-(2+\w)\xi_0  \cr
&\quad \qquad
+\smallfrac1{M_D}\[-(2+\w^2)\xmm+2\w(1+\w)\xp-(2+\w)(\Xone-2\Xthree)
      +2\w(1-\w)\Xtwo\] \cr
&\quad \qquad
+\smallfrac1{M_B}\[(2+\w^2)\xmm-2\w(1+\w)\xp-(2+\w)(\Xone-2\Xthree)
      +2\w(1-\w)\Xtwo\] \Bigr\} \cr
&-{\gB \gD \over 16 \pi^2 f^2}  C^{(1)}_2(1-\w^2)\Bigl\{\xi_0 \cr
&\quad \qquad
+\smallfrac1{M_D}\[ \w\xmm -2(1+\w)\xp+\Xone-2(1-\w)\Xtwo-2\Xthree\] \cr
&\quad \qquad 
+ \smallfrac1{M_B}\[-\w\xmm +2(1+\w)\xp+\Xone-2(1-\w)\Xtwo-2\Xthree\] 
      \Bigr\}  \cr }}
and
\eqn\hsix{\eqalign{
h_6^{B\to D} &= \(1 - {3 (g_B^2 D^{(1B)}_1
    + g_D^2 D^{(1D)}_1)\over 32 \pi^2 f^2}\) \Bigl\{\xo\cr
&\quad \qquad
+ \smallfrac1{M_D} \[-(2-\w)\xmm - 2(1+\w)\xp +\Xone +2 (1-\w)\Xtwo
    +6 \Xthree\] \cr
&\quad \qquad
+ \smallfrac1{M_B} \[(2-\w)\xmm + 2(1+\w)\xp +\Xone +2 (1-\w)\Xtwo
    +6 \Xthree\] \Bigr\} \cr
&{\gB \gD \over 16 \pi^2 f^2} C^{(1)}_1 \Bigl\{-(2+\w)\xi_0  \cr
&\quad \qquad
+\smallfrac1{M_D}\[(2+\w^2)\xmm-2\w(1+\w)\xp-(2+\w)(\Xone-2\Xthree)
      +2\w(1-\w)\Xtwo\] \cr
&\quad \qquad
+\smallfrac1{M_B}\[-(2+\w^2)\xmm+2\w(1+\w)\xp-(2+\w)(\Xone-2\Xthree)
      +2\w(1-\w)\Xtwo \] \Bigr\} \cr
&-{\gB \gD \over 16 \pi^2 f^2}  C^{(1)}_2(1-\w^2)\Bigl\{\xi_0 \cr
&\quad \qquad
+\smallfrac1{M_D}\[-\w\xmm +2(1+\w)\xp+\Xone-2(1-\w)\Xtwo-2\Xthree\] \cr
&\quad \qquad
+ \smallfrac1{M_B}\[\w\xmm -2(1+\w)\xp+\Xone-2(1-\w)\Xtwo-2\Xthree\]
      \Bigr\}  \cr }}
For decays to vectors, we have
\eqn\hfour{\eqalign{
h_4^{B\to D^*} &= \Dstring \Bigl\{ -\xo 
+\smallfrac1{M_D}\[ -\xmm -\cone +2\cthr \]\cr
&\qquad \quad
+\smallfrac1{M_B}\[-(2-\w)\xmm-2(1+\w)\xp-\cone-2(1-\w)\ctwo-6\cthr\]
         \Bigr\}\cr
&-\gbd C_1^{(5)} \Bigl\{ \xo 
+\smallfrac1{M_D}\[(2-\w)\xmm+2(1+\w)\xp+\cone+2(1-\w)\ctwo+6\cthr \]\cr
&\qquad \quad
+\smallfrac1{M_B}\[\xmm+\cone-2\cthr \]\Bigr\}\cr
&-\gbds C_1^{(3)}  \Bigl\{(1+\w) \xo \cr
&\qquad \quad
+\smallfrac1{M_D}\[2\w\xmm -2(1+\w)\xp +(1+\w)(\cone-2\cthr) 
                                           -2(1-\w)\ctwo \] \cr
&\qquad \quad
+\smallfrac1{M_B}\[(1+\w^2)\xmm -2\w(1+\w)\xp +(1+\w)(\cone-2\cthr) 
                                 -2\w(1-\w)\ctwo\] \Bigr\} \cr %
&-\gbds C_2^{(3)} (1-\w^2) \Bigl\{\xo 
+\smallfrac1{M_D}\[\xmm + \cone -2 \cthr\] \cr
&\qquad \quad
+\smallfrac1{M_B}\[\w \xmm -2 (1+\w) \xp +\cone -2(1-\w)\ctwo -2 \cthr \]
           \Bigr\} \cr }}
\eqn\hthree{\eqalign{
h_3^{B\to D^*} &= \Dstring \Bigl\{ -(1+\w)\xo \cr
&\qquad \quad
+\smallfrac1{M_D}\[(1-\w)\xmm -(1+\w)(\cone - 2\cthr) \]\cr
&\qquad \quad
+\smallfrac1{M_B}\[-(2-\w)(1-\w)\xmm -2(1-\w^2)(\xp+\ctwo) 
            -(1+\w)(\cone+6\cthr) \] \Bigr\} \cr
&-\gbd C_1^{(5)} \Bigl\{ (1+\w)\xo \cr
&\qquad \quad
+\smallfrac1{M_D}\[-(2-\w)(1-\w)\xmm-2(1-\w^2)(\xp-\ctwo) 
            +(1+\w)(\cone+6\cthr) \]\cr
&\qquad \quad
+\smallfrac1{M_B}\[-(1-\w)\xmm +(1+\w)(\cone - 2\cthr) \]
                      \Bigr\} \cr
&- \gbds C_1^{(3)} \Bigl\{ (1+\w)^2 \xo \cr
&\qquad \quad
+\smallfrac1{M_D}\[-2\w(1-\w)\xmm +2(1-\w^2)(\xp-\ctwo)
          +(1+\w)^2 (\cone - 2 \cthr) \] \cr
&\quad 
+\smallfrac1{M_B}\[-(1-\w)(1+\w^2)\xmm +2 \w (1-\w^2)(\xp-\ctwo)
  +(1+\w)^2  (\cone - 2 \cthr) \]\Bigr\}\cr
&- \gbds (1-\w^2) C_2^{(3)} \Bigl\{-(1+\w)\xo 
+\smallfrac1{M_D}\[(1-\w)\xmm-(1+\w)(\cone-2\cthr)\]\cr
&\qquad \quad
+\smallfrac1{M_B}\[\w(1-\w)\xmm -2 (1-\w^2) (\xp-\ctwo)
-(1+\w)(\cone-2\cthr)\] \Bigr\} \cr }}
\eqn\htwo{\eqalign{
h_2^{B\to D^*} &= \Dstring \Bigl\{ \xo
+\smallfrac1{M_D}\[2\xp +\cone -2 \ctwo -2 \cthr \] \cr
&\qquad \quad
+\smallfrac1{M_B}\[(2-\w)\xmm+2(1+\w)\xp+\cone+2(1-\w) \ctwo 
      +6\cthr\]\Bigr\}\cr
&-\gbd\w C_1^{(5)} \Bigl\{ \smallfrac1{M_B}\[\xmm -2 \xp -2\ctwo \]\Bigr\}\cr
&-\gbd C_2^{(5)} \Bigl\{ (1+\w) \xo \cr
&\qquad \quad
+\smallfrac1{M_D}\[-(2-\w)(1-\w)\xmm -2(1-\w)^2(\xp-\ctwo) 
           +(1+\w) (\cone +6 \cthr )\] \cr
&\qquad \quad
+\smallfrac1{M_B}\[-(1-\w)\xmm + (1+\w)(\cone -2  \cthr )\] \Bigr\} \cr
&-\gbds C_1^{(3)}\Bigl\{-(2+\w) \xo \cr
&\qquad \quad
+\smallfrac1{M_D}\[2(1-\w)\xmm +2\w (\xp-\ctwo) 
      -(2+\w)(\cone-2\cthr) \]\cr
&\qquad \quad
+\smallfrac1{M_B}\[-(2-\w+\w^2)\xmm +2\w^2 (\xp-\ctwo) 
                           -(2+\w)(\cone-2\cthr) \] \Bigr\} \cr
&-\gbds C_2^{(3)}\Big\{-\w(1+\w) \xo 
+\smallfrac1{M_D}\[\w(1-\w)\xmm -\w(1+\w)(\cone -2\cthr)  \]\cr
&\qquad \quad
+\smallfrac1{M_B}[-\smallfrac12(1+\w-3\w^2+\w^3)\xmm \cr
&\qquad\qquad\qquad\qquad
       +(1-\w)(1-\w^2)(\xp-\ctwo )   -\w(1+\w)(\cone-2\cthr) 
   ] \Big\} \cr }}
\eqn\hone{\eqalign{
h_1^{B\to D^*} &= \Dstring \Bigl\{\smallfrac1{M_D}\[-\xmm +2\xp +
             2 \ctwo\]\Bigr\}\cr 
&-\gbd \w C_1^{(5)} \Bigl\{ -\xo 
+\smallfrac1{M_D}\[(2-\w)\xmm + 2(1+\w)\xp -\cone -2(1-\w)\ctwo
              -6 \cthr \] \cr
&\qquad \quad
+\smallfrac1{M_B}\[ (1-\w)\xmm+2\w\xp-\cone-2\w\ctwo+2\cthr\]\Bigr\}\cr
&-\gbd C_2^{(5)} \Bigl\{-(1+\w) \xo \cr
&\qquad \quad
+\smallfrac1{M_D}\[-(2-3\w+\w^2)\xmm -2(1-\w^2)(\xp+\ctwo) 
         -(1+\w)(\cone+6\cthr) \]\cr
&\qquad \quad
+\smallfrac1{M_B}\[\w(1-\w)\xmm -2(1-\w^2)(\xp-\ctwo) 
         -(1+\w)(\cone-2 \cthr) \] \Bigr\} \cr 
&-\gbds C_1^{(3)}\Bigl\{ \xo 
+\smallfrac1{M_D}\[2\xp +\cone -2\ctwo -2\cthr \]\cr
&\qquad \quad
+\smallfrac1{M_B}\[-(1-\w)\xmm -2\w \xp +\cone +2\w \ctwo -2\cthr \]\Bigr\}\cr
&-\gbds C_2^{(3)}\Bigl\{(1+\w) \xo 
+\smallfrac1{M_D}\[-(1-\w)\xmm +(1+\w)(\cone -2\cthr) \] \cr
&\qquad \quad
+\smallfrac1{M_B}\[-\w(1-\w)\xmm +2(1-\w^2)(\xp-\ctwo)  
          +(1+\w)(\cone-2 \cthr) \] \Bigr\} \cr}}

Eqs.~\hfive--\hone\ constitute our main result. They express the form
factors for $B\to\{D,D^*\}$ in terms of five non-perturbative form
factors and the computable functions listed in  Appendix~B. The
corresponding form factors for $B_s\to\{D_s,D_s^*\}$, 
$h_i^{B_s\to D_s}$, may be obtained from those above by the
substitutions $C_i^{(m)} \to C_i^{(m+1)}$ and
$D_i^{(m)} \to D_i^{(m+1)}$. Our form factors agree with reference
\savjen\ in the $M \to \infty$ limit.

No analytic counterterms have been included in these formulas.
As one can see from eq. \jMm,
including analytic counterterms is equivalent to defining a strange
system Isgur-Wise function $\xi^{(s)}_0$ and new \ord\ correction
functions $\xi^{(s)}_{-}$, $\xi^{(s)}_{+}$, $\chi^{(s)}_{1}$,
$\chi^{(s)}_{2}$, $\chi^{(s)}_{3}$.  Thus, predictive power is lost.
The non-analytic corrections remain interesting however, because many
phenomenological models omit or improperly account for them (for
example, nonrelativistic quark models or quenched lattice
calculations\ref\golterman{M.F.L. Golterman, 
%``Chiral Perturbation Theory And The Quenched Approximation Of QCD,'' 
WASH-U-HEP-94-63, Oct
1994. (hep-lat/9411005)\semi C.W. Bernard and M.F.L. Golterman,
\physrev{D46}{1992}{853} (hep-lat/9204007)\semi C.W. Bernard and
M.F.L. Golterman, Nucl.\ Phys.\ B (Proc.Suppl.) 26 (1992) 360\semi
S. Sharpe, \physrev{D46}{1992}{3146}}), and may be improved by making
them consistent with the above formulas.  Along these lines, an
examination of the effect of quenching on heavy meson decay constants
has been performed in reference \ref\booth{M. Booth, preprint
IFT-94-10, hep-ph/9412228}. The present work allows a similar analysis
to be done for semileptonic $B \to D$. The theoretical relation
between strange and non-strange form factors is especially important
for lattice computations, which typically extrapolate non-strange form
factors from simulations involving strange quarks\latt. 

\newsec{ Luke's Theorem}

We may check our calculation by verifying consistency with Luke's Theorem,
which says $\frac1M$ corrections to our hadronic matrix elements must 
vanish at threshold. It is easy to see that this holds true for 
$\bra{D(v)} \cgb \ket{B(v)} $, but the decay to $D^*$ is less transparent.

The subleading Isgur-Wise functions drop out of the only surviving
$B \to D^*$ form factor $h_3$ at threshold because 
the coefficients of $\lbar \xo, \xp$, and $\Xtwo$ are proportional to
$1-\w$, while the functions $\Xone$ and $\Xthree$  vanish identically
at $\w =1$. The sum of $\ol B \to D^*$ graphs is thus
\eqn\bdsthres{\eqalign{
&\bra{D^*(\eps_D,v)} \cgb \ket{B(v)} = -2 \eps_D^{*\lambda}\cr
& - \eps_D^{*\lambda} {1  \over 16 \pi^2 f_K^2}
\Biggl[2 \gB \gD C_1^{(5)}\Big|_{\w=1} + 4 \gB \gDs C_1^{(3)}\Big|_{\w=1}
 -2 g_{D^*}^2 D_1^{(5D)} - g_D^2 D_1^{(3D)}
 -3 g_B^2 D_1^{(1B)} \Biggr] \cr }}
Luke's theorem requires that the term in brackets vanish to \ord~.
The $\CO(1)$ terms vanish trivially because the various integrals
$C_1^{(i)}$, $D_1^{(i)}$ are equal to each other at $\w =1$ to
\ord~. There are two contributions at \ordMm . The \ord\ pieces of
the axial couplings can multiply the $\CO(m_s)$ parts of the integrals
to give a sum proportional to 
$$ 2 g_B g_D + 4 g_B g_{D^*} -2 g_{D^*}^2 -g_D^2 -3 g_B^2 =0 +
      \CO(\frac1{M_D^2},\frac1{M_B^2},\frac1{M_B M_D}), $$
or the \ordMm\ parts of the integrals can multiply the 
$\CO(1)$ axial couplings to give a sum proportional to
$$2 C_1^{(5)} + 4 C_1^{(3)} - 2 D_1^{(5D)} - D_1^{(3D)}
              -3 D_1^{(1D)}. $$
By writing $a = \frac\delta{m}( \cos \Phi + \sin \Phi) + x $ and
expanding the integrals to linear order in $x$,
one can show that this sum is proportional to
$$\int d\Phi K(b,\delta( \cos \Phi + \sin \Phi) ) (3\Delta^B + \Delta^D)
(  \cos \Phi - \sin \Phi ) = 0, $$
where $K$ is a function even under interchange of $\cos \Phi$ and
$\sin \Phi$.

\newsec{ Discusion}

Previous analyses\chow\ included only $\frac1M$ corrections due to $D$ meson
hyperfine splitting. Here, we also include $B$ meson hyperfine splittings,
\ord\ axial coupling corrections, and \ord\ corrections to the current.

We identify contributions which are readily separated from contact terms by 
their parametric behavior in the $m \to 0$ and $\Delta \to 0$ limits.  
This includes both
chiral logs which go like $m^2 \ln \frac{\mu^2}{m^2}$ and functions 
which depend on the ratio ${\Delta \over m }$.  Such behavior can
never arise from contact terms because the counterterms are proportional
to positive, integer powers of the light quark masses.

The nonanalytic dependence of the form factors in eqs. \hfive\ to
\hone\ arises from the integrals $C_i$ in eqs. (B.2)  and (B.3) of 
Appendix B. These integrals are finite in the $m \to 0$ and 
$\Delta, \Delta' \to 0$ limits, and may be expanded about $m=0$ or
$\Delta =0$ after choosing the behavior of the ratio ${\Delta \over m}$
in the double limit. 
%It was pointed out in reference \chow\ that 
At a fixed order in $\frac1M$, terms in $C_i$ which go like
$({\Delta  \over m})^n $ dominate the heavy quark
chiral expansion\chow\ if $\Delta$ is held fixed as $m \to 0$ (terms 
which go like $\Delta^2 \ln m^2$ cancel). However, if one
does not assume this behavior of ${\Delta \over m}$ (for example,
one might instead hold the ratio fixed at a physical value), other
contributions in the heavy quark chiral expansion (such as axial 
corrections $g_i$ and current corrections $\rho_i$) are equally or
more important.

Because the functions $C_i$ are proportional to at least two powers of
$\Delta$ or $\Delta'$, the $\CO(\frac1M, m_s)$ corrections in
eqs.~\hfive\ to
\hone\ are proportional to either axial couplings $g_1, g_2$ or
subleading Isgur-Wise functions $\xmm, \xp, \Xone, \Xtwo, \Xthree$.
It is only at $\CO(\frac1{M^2})$  or $\CO(\frac1M, m_s^2)$
that hyperfine splittings also enter. 

In principle, our results can be used to estimate $B_s \to D_s l \nu$
form factors once $B \to D l \nu$ form factors are measured. The 
necessary chiral lagrangian parameters $g, g_1$ and $g_2$ can be extracted
from processes such as $B \to \pi l \nu$ and $B \to l \nu$\US,
$D^* \to D \pi$\wbdy, and $D^* \to D \gamma$\ref\amund{J. Amundson, \etal,
\pl{296}{1992}{415}\semi
H-Y Cheng et al, \physrev{D47}{1993}{1030}\semi
P. Cho and H. Georgi, \pl{B296}{1992}{408}; erratum ibid {\bf
B300}(1993)410}.  For a more reliable estimate, analytic counterterms should
be included (from a model or lattice computation).

To see the effect of the \ord\ and SU(3) corrections, we
examine a quantity which is sensitive only to simultaneous violations 
of both symmetries, the ratio $ R(\w) = { h_5^{B_s \to D_s} / h_6^{B_s \to D_s}
\over h_5^{B \to D} / h_6^{B \to D} }$. In general, such 
ratios of form factors will depend on all the chiral parameters and subleading
Isgur-Wise functions, but because of constraints due to Luke's theorem, this
particular ratio takes a simple form,
\eqn\ratio{\eqalign{ 
R(\w) -1 &= 
{- g^2 \over 16 \pi^2 f^2}\(\frac{\lbar}{M_D} - \frac{\lbar}{M_B}\)
 \Biggl\{ \Bigl[ 3(2-\w)\( D_1^{(2B)} - D_1^{(1B)} + D_1^{(2D)} - D_1^{(1D)} \)
                  \cr
           &  -2(2+ \w^2)\( C_1^{(2)} - C_1^{(1)} \) 
            + 2 \w( 1 - \w^2) \(C_2^{(2)} - C_2^{(1)}\)\Bigr] \cr
    &+ {2 \xp \over \lbar \xo }(1 + \w) \Bigl[
    3 \( D_1^{(2B)} - D_1^{(1B)} + D_1^{(2D)} - D_1^{(1D)} \) \cr
    &+ 2 \w \( C_1^{(2)} - C_1^{(1)} \)
    - 2 (1 -\w^2) \( C_2^{(2)} - C_2^{(1)} \) \Bigr] \Biggr\}. \cr }}
We retain the full dependence on the ratio ${\Delta \over m}$, but
dependence on $g_1, g_2, \Xone, \Xtwo$ and $\Xthree$  drop out, 
to leading order in $\frac1M$.

\nref\lnn{Z. Ligeti, Y. Nir, and M. Neubert,
\physrev{D49}{1994}{1302} }

\INSERTFIG{fig_r.eps}{3. }{The quantity $R(\w)-1 \equiv 
{ h_5^{B_s \to D_s} / h_6^{B_s \to D_s}
\over h_5^{B \to D} / h_6^{B \to D} }-1$ as given by eqn.~\ratio, with $g^2 =
0.5$, $\mu=1.0\gev$ and  $\lbar = 0.5 \gev$.  For the solid line we
used $(1 +
\w){2 \xp \over \lbar \xo } = \w - 2.2 \pm 0.4 $ from  a QCD sum rule
computation\lnn. For the dashed line we used $\xp=0$. }

For a rough idea of the size and shape of $R(\w)$, we take $g^2 =
0.5$, $\mu=1.0\gev$, and use the results of a QCD sum rule
computation\lnn\ as input: $\lbar = 0.5 \gev$ and $(1 +
\w){2 \xp \over \lbar \xo } = \w - 2.2 \pm 0.4 $.  The solid line in
\fig\rrplot{The quantity $R(\w)-1 \equiv 
{ h_5^{B_s \to D_s} / h_6^{B_s \to D_s}
\over h_5^{B \to D} / h_6^{B \to D} }-1$ as given by eqn.~\ratio, with $g^2 =
0.5$, $\mu=1.0\gev$ and  $\lbar = 0.5 \gev$.  For the solid line we
used $(1 +
\w){2 \xp \over \lbar \xo } = \w - 2.2 \pm 0.4 $ from  a QCD sum rule
computation\lnn. For the dashed line we used $\xp=0$.}\
 shows $R(\w)-1$ for these values while the dashed line shows it
for $\xp=0$. 

The surprisingly large deviation from unity arises because every term
inside the parentheses of eq. \ratio\ adds constructively. Special
values of ${\xp \over \lbar \xo }$ may give smaller deviations from
unity, but fig.~3 represents the typical scale of symmetry breaking
for this quantity. In order to estimate the size of the counterterms
we study $R(\w)-1$ as we increase $\mu$. For $\mu=2\gev$ the dashed
curve in fig.~3 is virtually unchanged but the solid curve nearly
doubles. Such a large effect casts doubt on the validity of the
simultaneous heavy quark and SU(3) chiral symmetry expansion for this
system. Even with $\lbar$ as small as $0.25\gev$, the roughly $15\%$
corrections to $R(\w) -1$ are much larger than expected. Thus, care
should be taken when extrapolating form factors from the strange to
non-strange $B \to D l\nu$ systems, especially if, as in
lattice calculations\latt, both heavy quark and chiral 
extrapolations are performed.

\newsec{ Conclusions}
We have computed the \ordMm\ heavy quark and SU(3) corrections to $B_s
\to D_s e \nu$ form factors.  If the analytic counterterms are
neglected, $B_s \to D_s e \nu$ form factors are given in terms of the
the $B \to D e \nu$ form factors, the leading order chiral
parameter $g$, and two \ord\ chiral parameters $g_1$ and $g_2$.
All the chiral parameters can be extracted, in principle, from other
heavy meson decays, so all six potentially observable $B_s \to D_s l
\nu$ form factors are determined, in the formal chiral limit, by the
six form factors of $B \to D l \nu$.

We say ``formal'' because, while the non-analytic terms are the leading
terms in the chiral expansion for small kaon mass, in reality one
expects analytic counterterms to be numerically just as important.
Analytic counterterms proportional to the strange quark mass have also
been presented, but no predictive power remains when they are
included. The chiral log corrections remain interesting however,
because many phenomenological models (such as  nonrelativistic quark models
or quenched lattice calculations) omit or improperly account for
the chiral log contribution and may be improved by making them
consistent with the above formulas. In addition, when such loop corrections
are large, they warn of a possible breakdown of the chiral expansion. 

Precisely this situation occurs for the quantity  
$ R(\w) = { h_5^{B_s \to D_s} / h_6^{B_s \to D_s}
\over h_5^{B \to D} / h_6^{B \to D} }$, where we find typical 
deviations from the symmetry limit of $15\%$ to $30\%$. This is
alarming, since these deviations involve simultaneous heavy quark and chiral
symmetry violations, and are therefore expected to be only a few percent.
Care should be taken when using heavy quark and SU(3) symmetry relations 
for the extraction of $B \to D l \nu$ form factors from those of
$B_s \to D_s l \nu$.

\bigskip
\qquad  ACKNOWLEDGMENTS
\bigskip

B.G. is supported in part by the Alfred P.~Sloan Foundation. G.B. and
B.G. acknowledge the support of the Department of Energy, under
contract DOE--FG03--90ER40546.

\vfill\eject

\appendix{A}{Feynman Rules for the Current}
The Feynman rules for the current may be conveniently summarized by
expressing the form factors in terms of the current parameters. Tree level, 
$\CO(\frac1{M_D},m_s^0)$ values for these form factors are given in table 
1 (for example, $f_1 = \xo + \frac1{M_D}[\r1 + 2 \r3 -\r5 -2 \r6] +
\CO(\frac1{M_B})$). 
\def\R#1{\rho_#1$\hfil$\null}
\bigskip
\vbox{\centerline{Table 1.}\medskip
\hfil\vbox{\offinterlineskip
\hrule
\halign{&\vrule#&\strut\quad\hfil$#$\quad\cr
height2pt&\omit&&\omit&&\omit&&\omit&&\omit&&\omit&\cr
& &&\R1&&\R3&&\R4&&\R5&&\R6&\cr
height3pt&\omit&&\omit&&\omit&&\omit&&\omit&&\omit&\cr
\noalign{\hrule}
height2pt&\omit&&\omit&&\omit&&\omit&&\omit&&\omit&\cr
&f_0&&0 &&0 &&0 &&0 &&0 &\cr
&f_1&&1 &&2 &&0 &&-1 &&-2 &\cr
&f_2&&-1 &&-2 &&0 &&1 &&2 &\cr
&f_3&&1+\w &&-4+2\w &&12 &&1+\w &&-4+2\w &\cr
&f_4&&1 &&0 &&-4 &&-1 &&0 &\cr
&f_5&&1 &&-2 &&0 &&1+2\w &&-2 &\cr
&f_6&&0 &&0 &&0 &&-2 &&0 &\cr
&f_7&&0 &&0 &&0 &&0 &&0 &\cr
&f_8&&-1 &&2 &&0 &&1 &&-2 &\cr
&f_9&&-1 &&0 &&4 &&1 &&0 &\cr
&f_{10}&&-1 &&0 &&4 &&-1 &&0 &\cr
&f_{11}&&-1 &&2 &&0 &&-1-\w &&2 &\cr
&f_{12}&&0 &&0 &&0 &&0 &&0 &\cr
&f_{13}&&0 &&0 &&0 &&1 &&0 &\cr
&h_1&&0 &&0 &&0 &&-2 &&4 &\cr
&h_2&&-1 &&0 &&4 &&-1 &&0 &\cr
&h_3&&1+\w &&-2 &&-4\w &&1+\w &&-2 &\cr
&h_4&&1 &&0 &&-4 &&-1 &&0 &\cr
&h_5&&-1 &&-2 &&0 &&1 &&2 &\cr
&h_6&&-1 &&4 &&-12 &&-1-2\w &&4-4\w &\cr
}
\hrule}
\hfil}
\bigskip

Heavy quark symmetry gives
the $\CO(\frac1{M_B})$ parts $f_i^B,h_i^B$, of these form factors 
simply in terms of the $\CO(\frac1{M_D})$ parts $f_i^D,h_i^D$, by
the following relations:
\eqn\bparts{
 \eqalign{ f_0^B &= -h_1^D\cr f_5^B &= f_4^D \cr f_{10}^B &= f_{11}^D
\cr h_2^B &= h_5^D \cr}
\qquad 
 \eqalign{f_1^B &= h_4^D \cr f_6^B &= 0 \cr   f_{11}^B &= f_{10}^D\cr
 h_3^B &= f_3^D \cr}
 \qquad 
 \eqalign{f_2^B &= h_2^D \cr  f_7^B &= f_6^D \cr  f_{12}^B &= f_{13}^D\cr 
 h_4^B &= f_1^D\cr}
\qquad 
 \eqalign{f_3^B &= h_3^D \cr f_8^B &= f_9^D  \cr  f_{13}^B &=  f_{12}^D\cr
 h_5^B&= h_6^D \cr}
\qquad
 \eqalign{f_4^B &= h_5^D  \cr f_9^B &= f_8^D  \cr h_1^B &= 0 \cr
h_6^B &= h_5^D \cr}}

\appendix{B}{Integral and definitions}
The one loop corrections involve the integral
\eqn\Cdef{\eqalign{
  C^{\alpha\beta}(\w,m,\Delta,\Delta') &= \kintegral { k^\alpha k^\beta
   \over  (k^2 - m^2)(\vk - \Delta)(\vpk - \Delta')  } \cr
    &= {\im \over 16 \pi^2} [C_1(\w,m,\Delta,\Delta') g^{\alpha\beta} 
 + C_2(\w,m,\Delta,\Delta') ( v^\alpha v'^\beta + v^\beta v'^\alpha) ]\cr
 &+ C_3(\w,m,\Delta,\Delta')  v'^\alpha v'^\beta
    + C_4(\w,m,\Delta,\Delta') v^\alpha v^\beta \cr
                             } }

The integrals $C_1 - C_4$ may be expressed as one dimensional integrals
\eqn\Integ{\eqalign{
  C_1 &= m^2 \int_0^\frac\pi2 d \Phi \biggl[
   (\frac2\epsilon + \ln 4\pi-\gamma_E +1)(2 a^2 b^2 - b)+2 a^2 b^2 \cr
   &+  (b - 2 a^2 b^2)\log
       -  4 a b \sqrt{a^2 b^2 -b}  \ln[a \sqrt{b} + \sqrt{a^2 b-1}] \biggr]\cr  }}
\eqn\Integr{\eqalign{
%  C_{\pmatrix{2\cr3\cr4\cr}} &=  -2 m^2 \int_0^{\pi \over 2} d\Phi 
  C_i &=  -2  m^2 \int_0^{\pi \over 2} d\Phi 
   \(\alpha_i(\Phi)\) \biggl[- b^2 (1 - 4 a^2 b)(\frac2\epsilon 
    + \ln 4\pi-\gamma_E  +1) \cr
   &+  b^2 (1 - 4 a^2 b) \log 
     + b^2(1 -5 a^2 b)
        +{ 2 a b^3 (3 - 4 a^2 b) \over \sqrt{a^2 b^2 -b} } 
    \ln[a \sqrt{b} + \sqrt{a^2 b-1}] \biggr] \cr }}
where $a = \frac\Delta{m} \cos \Phi + \frac{\Delta'}{m} \sin \Phi$, 
$b = (1 + 2 \w \cos\Phi \sin\Phi)^{-1}$, and $\alpha_2(\Phi) = \cos \Phi \sin \Phi$,
$\alpha_3(\Phi) = \sin^2 \Phi$, $\alpha_4(\Phi) = \cos^2 \Phi$.

We will work in a scheme such that $\frac2\epsilon + \ln 4\pi-\gamma_E  +1=0$. 
In this scheme, for $\Delta = \Delta' =0$, these integrals simplify to
\eqn\catzero{\eqalign{
C_1 &=  m^2 \log r(\w), \cr
C_2 &= -  m^2 {(\log +1)\over 1-\w^2}[1 - \w r(\w)], \cr
C_3 &= C_4 = - m^2 {(\log +1)\over 1-\w^2}[r(\w) -\w ] \cr }}
\OMIT{
\eqn\catzero{\eqalign{
C_1 &= \im m^2 \log {\ln(\w + \sqrt{\w^2 -1})\over \sqrt{\w^2 -1}}, \cr
C_2 &= - \im m^2 {(\log +2)\over 1-\w^2}
   \left[1- \frac{\w \ln(\w + \sqrt{\w^2 -1})}{\sqrt{\w^2-1}}\right]. \cr 
C_3 &= C_4 = -\im m^2 \left(\log +2\right) \left[\frac{\w}{\w^2-1} + 
            {\ln(\w + \sqrt{\w^2 -1})\over (1-\w^2)\sqrt{\w^2 -1}}\right],
          \cr }}
}
where $r(\w) = {\ln(\w + \sqrt{\w^2 -1}) \over \sqrt{\w^2 -1} }$
and $r(\w) \to 1$ at threshold.

The following linear combinations of the integrals $C_i$ will be useful:
\eqn\Ccomb{\eqalign{
  \C{i}1 &= C_i(\w,m_K,\Delta^{(B)}+\delta,\Delta^{(D)}+\delta)\cr
        &\qquad\qquad+  \smallfrac32 C_i(\w,m_\pi,\Delta^{(B)},\Delta^{(D)})
        + \smallfrac16 C_i(\w,m_\eta,\Delta^{(B)},\Delta^{(D)})\cr
   \C{i}2 &= 2 C_i(\w,m_K,\Delta^{(B)}-\delta,\Delta^{(D)}-\delta)
       + \smallfrac23  C_i(\w,m_\eta,\Delta^{(B)},\Delta^{(D)})\cr
  \C{i}3 &= C_i(\w,m_K,\Delta^{(B)}+\delta,\delta) 
        + \smallfrac32 C_i(\w,m_\pi,\Delta^{(B)},0)
        +\smallfrac16 C_i(\w,m_\eta,\Delta^{(B)},0)\cr
   \C{i}4 &=  2 C_i(\w,m_K,\Delta^{(B)}-\delta,-\delta) 
          + \smallfrac23  C_i(\w,m_\eta,\Delta^{(B)},0) \cr
  \C{i}5 &=  C_i(\w,m_K,\Delta^{(B)}+\delta,-\Delta^{(D)}+\delta)\cr
        &\qquad\qquad+ \smallfrac32 C_i(\w,m_\pi,\Delta^{(B)},-\Delta^{(D)})
        + \smallfrac16 C_i(\w,m_\eta,\Delta^{(B)},-\Delta^{(D)})\cr
   \C{i}6 &=  2 C_i(\w,m_K,\Delta^{(B)}-\delta,-\Delta^{(D)}-\delta)
          + \smallfrac23  C_i(\w,m_\eta,\Delta^{(B)},-\Delta^{(D)})\cr }}
are produced when summing over intermediate states contributing to
vertex corrections, while
\eqn\Dcomb{\eqalign{
  \D{i}{1B} &= C_i(1,m_K,\Delta^{(B)}+\delta,\Delta^{(B)}+\delta)\cr
       &\qquad\qquad +  \smallfrac32 C_i(1,m_\pi,\Delta^{(B)},\Delta^{(B)})
        + \smallfrac16 C_i(1,m_\eta,\Delta^{(B)},\Delta^{(B)})\cr
   \D{i}{2B} &= 2 C_i(1,m_K,\Delta^{(B)}-\delta,\Delta^{(B)}-\delta)
       + \smallfrac23  C_i(1,m_\eta,\Delta^{(B)},\Delta^{(B)})\cr
  \D{i}{3B} &= C_i(1,m_K,-\Delta^{(B)}+\delta,-\Delta^{(B)}+\delta) \cr
         &\qquad\qquad+ \smallfrac32 C_i(1,m_\pi,-\Delta^{(B)},-\Delta^{(B)})
        +\smallfrac16 C_i(1,m_\eta,-\Delta^{(B)},-\Delta^{(B)})\cr
   \D{i}{4B} &=  2 C_i(1,m_K,-\Delta^{(B)}-\delta,-\Delta^{(B)}-\delta) 
          + \smallfrac23  C_i(1,m_\eta,-\Delta^{(B)},-\Delta^{(B)}) \cr
  \D{i}{5B} &=  C_i(1,m_K,\delta,\delta)
        + \smallfrac32 C_i(1,m_\pi,0,0)
        + \smallfrac16 C_i(1,m_\eta,0,0)\cr
   \D{i}{6B} &=  2 C_i(1,m_K,-\delta,-\delta) 
       + \smallfrac23  C_i(\w,m_\eta,0,0)\cr }}
and the analogous integrals, $D_{i}^{(jD)}$ with $\Delta^{(B)} \to
\Delta^{(D)}$, arise from wavefunction renormalization.

\def\CC#1{$C_1^{(#1)}$}
\nfig\plfour{The real part of the linear combinations of the 
integral $C_1$ defined in eq.~\Ccomb, as a function of $\w$. 
The integral was done numerically for 
physical values of the masses.}
\def\CC#1{$C_2^{(#1)}$}
\nfig\plfive{The  real part of the linear combinations of the 
integral $C_2$ defined in eq.~\Ccomb, as a function of $\w$. 
The integral was done numerically for 
physical values of the masses.}
\def\CC#1{$C_3^{(#1)}$}
\nfig\plsix{The  real part of the linear combinations of the 
integral $C_3$ defined in eq.~\Ccomb, as a function of $\w$. 
The integral was done numerically for 
physical values of the masses.}
The real parts of the functions $C_1^{(i)}(\w)$, $C_2^{(i)}(\w)$ and
$C_3^{(i)}(\w)$, for physical values of masses and mass splittings,
are plotted in figures \xfig\plfour, \xfig\plfive\ and \xfig\plsix, 
respectively. Only $C_i^{(5)}(\w)$
have non-vanishing imaginary parts, as expected from the corresponding
marginally allowed decay $D^* \to D \pi$.  Of these, only
$C_3^{(5)}(\w)$ has an appreciable imaginary part, which ranges from
$-0.06$ at $\w=1$ to $-0.04$ at $\w=1.6$.  Numerical values for the
real parts of $D_{i}^{(jB)}$ and $D_{i}^{(jD)}$ are given in Tables~2
and~3. Only $D_i^{(3D)}$ have nonzero imaginary parts, with values
$0.11$, $-0.40$ and $-0.75$ for $i=1,2,3$, respectively.
\vfill\eject

\ifx\answfig\yesanswfig%
\def\INSERTFIG#1#2#3{\vbox{\vbox{\epsfbox{#1}\hfill}%
{\narrower\noindent%
\multiply\baselineskip by 3%
\divide\baselineskip by 4%
\vskip0.5in{\ninerm Figure #2 }{\ninesl #3 \medskip}}
}}%

\def\CC#1{$C_1^{(#1)}$}

\INSERTFIG{fig_c1.eps}{4. }{The real part of the linear combinations of the 
integral $C_1$ defined in eq.~\Ccomb, as a function of $\w$. 
The integral was done numerically for 
physical values of the masses.}
\vskip-5.05in\hskip1.15in {\CC1} \hskip2.3in {\CC2}
\vskip1.6in\hskip1.15in {\CC3} \hskip2.3in {\CC4}
\vskip1.6in\hskip1.15in {\CC5} \hskip2.3in {\CC6}
\def\CC#1{$C_2^{(#1)}$}
\vfill\eject

\def\CC#1{$C_2^{(#1)}$}

\INSERTFIG{fig_c2.eps}{5. }{The  real part of the linear combinations of the 
integral $C_2$ defined in eq.~\Ccomb, as a function of $\w$. 
The integral was done numerically for 
physical values of the masses.}
\vskip-5.05in\hskip1.15in {\CC1} \hskip2.3in {\CC2}
\vskip1.6in\hskip1.15in {\CC3} \hskip2.3in {\CC4}
\vskip1.6in\hskip1.15in {\CC5} \hskip2.3in {\CC6}
\def\CC#1{$C_3^{(#1)}$}
\vfill\eject

\def\CC#1{$C_3^{(#1)}$}

\INSERTFIG{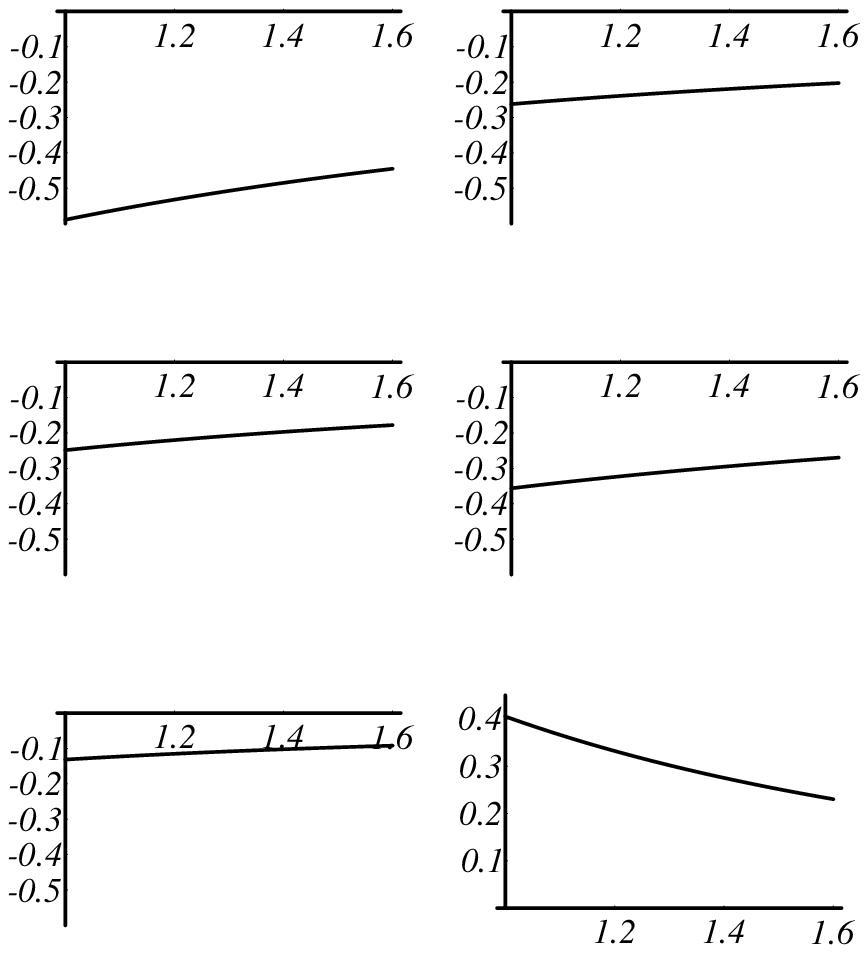}{6. }{The  real part of the linear combinations of the 
integral $C_3$ defined in eq.~\Ccomb, as a function of $\w$. 
The integral was done numerically for 
physical values of the masses.}
\vskip-5.05in\hskip1.15in {\CC1} \hskip2.3in {\CC2}
\vskip1.6in\hskip1.15in {\CC3} \hskip2.3in {\CC4}
\vskip1.7in\hskip1.15in {\CC5} \hskip2.3in {\CC6}
\vfill\eject
\fi
 
\def\DB#1{{D_i^{(#1B)}}}
\def\DD#1{{D_i^{(#1D)}}}
\bigskip
\vbox{\centerline{Table 2.}\medskip
\vbox{\offinterlineskip
\hrule
\halign{&\vrule#&\strut\quad\hfil$#$\quad\cr
height2pt&\omit&&\omit&&\omit&&\omit&&\omit&&\omit&&\omit&\cr
&i&&\DB1 &&\DB2&&\DB3&&\DB4&&\DB5&&\DB6&\cr
height3pt&\omit&&\omit&&\omit&&\omit&&\omit&&\omit&\cr
\noalign{\hrule}
height2pt&\omit&&\omit&&\omit&&\omit&&\omit&&\omit&\cr
&1&& 0.139&&-0.406&&-0.264&&-0.517&&-0.142&&-0.331&\cr
&2&& -0.216&&-0.177&&-0.055&&-0.034&&-0.107&&-0.184&\cr
&3&& -0.396&&-0.319&&-0.096&&-0.054&&-0.191&&-0.334&\cr
&4&& -0.396&&-0.319&&-0.096&&-0.054&&-0.191&&-0.334&\cr
}
\hrule}
\hfil}
\bigskip
\bigskip
\vbox{\centerline{Table 3.}\medskip
\vbox{\offinterlineskip
\hrule
\halign{&\vrule#&\strut\quad\hfil$#$\quad\cr
height2pt&\omit&&\omit&&\omit&&\omit&&\omit&&\omit&&\omit&\cr
&i&&\DD1 &&\DD2&&\DD3&&\DD4&&\DD5&&\DD6&\cr
height3pt&\omit&&\omit&&\omit&&\omit&&\omit&&\omit&\cr
\noalign{\hrule}
height2pt&\omit&&\omit&&\omit&&\omit&&\omit&&\omit&\cr
&1&& 0.696&&-0.223&&-0.021&&-1.144&&-0.142&&-0.331&\cr
&2&&-0.325&&-0.246&&-0.109&&0.440&&-0.107&&-0.184&\cr
&3&&-0.614&&-0.448&&-0.209&& 0.829&&-0.191&&-0.334&\cr
&4&&-0.614&&-0.448&&-0.209&& 0.829&&-0.191&&-0.334&\cr
}
\hrule}
\hfil}
\vfill\eject

\ifx\answfig\yesanswfig%
\listrefs
\else
\listrefs
\listfigs
\fi

\bye